\documentclass[%
aip,jap,
amsmath,amssymb,
reprint,
]{revtex4-2}

\usepackage{graphicx}
\usepackage{dcolumn}
\usepackage{bm}
\usepackage{amsthm}
\usepackage{color}
\usepackage[latin1,utf8]{inputenc}
\usepackage[T1]{fontenc}
\usepackage{textcomp}
\usepackage{mathptmx}
\usepackage{mathtools}
\usepackage{upgreek}

\newtheoremstyle{query}%
{}{}
{\color{red}}
{}
{\sffamily\bfseries}{:}{12pt}
{}
\theoremstyle{query}

\begin{document}
\title{Three-dimensional modeling of minority-carrier lateral diffusion length including random alloy fluctuations in (In,Ga)N and (Al,Ga)N single quantum wells}
\author{Huan-Ting Shen}
\affiliation{ 
Graduate Institute of Photonics and Optoelectronics and Department of Electrical Engineering, National Taiwan University, Taipei 10617, Taiwan
}%
\author{Claude Weisbuch}
\affiliation{ 
Materials Department, University of California, Santa Barbara, CA 93106-5050, USA
}%
\affiliation{ 
Laboratoire de Physique de la Mati$\grave{e}$re Condens$\acute{e}$e, Ecole Polytechnique, CNRS, IP Paris, 91128 Palaiseau Cedex, France
}

\author{James S. Speck}
\affiliation{ 
Materials Department, University of California, Santa Barbara, CA 93106-5050, USA
}%
\author{Yuh-Renn Wu} 
\email{yrwu@ntu.edu.tw}
\affiliation{ 
Graduate Institute of Photonics and Optoelectronics and Department of Electrical Engineering, National Taiwan University, Taipei 10617, Taiwan
}%
\date{\today}
\begin{abstract}
For nitride-based InGaN and AlGaN quantum well (QW) LEDs, the potential fluctuations caused by natural alloy disorders limit the lateral intra-QW carrier diffusion length and current spreading. The diffusion length mainly impacts the overall LED efficiency through sidewall nonradiative recombination, especially for $\mu$LEDs. In this paper, we study the carrier lateral diffusion length for nitride-based green, blue, and ultraviolet C (UVC) QWs in three dimensions. We solve the Poisson and drift-diffusion equations in the framework of localization landscape theory. The full three-dimensional model includes the effects of random alloy composition fluctuations and electric fields in the QWs. The dependence of the minority carrier diffusion length on the majority carrier density is studied with a full three-dimensional model. The results show that the diffusion length is limited by the potential fluctuations and the recombination rate, the latter being controlled by the spontaneous and piezo electric fields in the QWs and by the screening of the the by the screening of the internal electric fields by carriers.
\end{abstract}
\maketitle
%
%
\section{\label{sec:level1}Introduction}
The increasing demand for lighting, sterilization, and display units has led to traditional light sources being replaced by light-emitting diodes (LEDs), which offer high power efficiency. In LEDs, the diffusion length in the lateral direction and the defect density both impact the internal quantum efficiency and the possible current spreading, particularly as chip sizes decrease. The efficiency of nitride-based LEDs is limited in part by the density of nonradiative recombination (NR) centers. The natural random alloy fluctuations in InGaN are believed to cause carrier localization\cite{chichibu1997luminescences} and suppress the carrier diffusion to NR centers. However, the influence of random alloy potentials is difficult to analyze quantitatively due to computational limitations. Our recent studies with a full three-dimensional (3D) simulation model have investigated the vertical carrier transport with respect to alloy fluctuations and thickness fluctuations,\cite{yang2014influence,butte2018optical,weisbuch2020disorder} even in ultraviolet C (UVC) LEDs,\cite{sohi2018alloy} in which fluctuating potentials occur in both quantum wells (QWs) and quantum barriers (QBs). Hence, it is important to further investigate the impact of random alloy fluctuations on the lateral carrier transport.

The existence of random alloy fluctuations has been extensively studied. Usually, optical methods can resolve alloy fluctuations at scales down to around 50 nm,\cite{de2011two,mensi2018direct,sonderegger2006high,maruo1997three} mainly limited by the optical resolution. A recent scanning tunneling luminescence (STL) study directly observed localized states induced by the intrinsic disorder on a scale of a few nanometers.\cite{hahn2018evidence} Models including localized carrier effects in disordered systems indicate that these fluctuations change the carrier mobility and lifetime.\cite{sohi2018alloy,david2010droop,feix2017ga} In addition, the forward voltage V$_{F}$ is improved in LED structures with a strongly fluctuating potential, decreasing the effective barrier height for electron and hole injection into the QWs and increasing the radiative efficiency.\cite{nath2013unipolar,browne2015electron,qwah2020theoretical,nakamura1995high,nakamura2013history} Therefore, the natural alloy disorder that causes potential fluctuations in materials plays an important role in limiting carrier diffusion, preventing them from being captured by defects or surface states. When considering carrier diffusion in QW systems, the issues are more complicated.\cite{shklovskii2013electronic} When a QW is formed in a nitride-based device, the strain-induced piezoelectric polarization charge\cite{romanov2006strain} and spontaneous polarization charge will lead to net polar charges at the interfaces, which results in internal electric fields in the QWs and reduced radiative recombination rates due to decreased electron--hole wave function overlap. This also significantly affects carrier diffusion because the diffusion length depends on the diffusion coefficient and carrier lifetime, which is inversely proportional to the total recombination rate.

The impact of the diffusion length is most critical for $\mu$LEDs, as with chip size decreasing, minority carriers have a greater chance of diffusing to the sidewall. High-density surface states become major NR  centers. This effect is more severe in AlGaInP-based red LEDs than in nitride-based LEDs. As the carrier diffusion length determines the likelihood of carriers being trapped by surface states, the factors that influence the diffusion length in nitride-based LEDs should be investigated. Although we expect the potential fluctuations $\Delta$V to be important, our studies show that this phenomenon may be affected by additional factors. For example, according to the continuity equation, if the carriers' local radiative recombination rate is less than the injected current density, the excess carriers will be delocalized and leave the local confining potential regions within the QWs, effectively increasing the diffusion length.

This work investigates the impact of random alloy fluctuations on the lateral diffusion within nitride alloy QWs. Simulations play a vital role in understanding LED behavior and help guide improved designs. However, addressing the alloy fluctuation effects in an entire LED structure is a difficult simulation task requiring 3D simulation software. Alloy fluctuations cannot be described by one-dimensional (1D) simulation tools, which do not take in-plane fluctuations into account.\cite{karpov2011modeling,kisin2015inhomogeneous,geng2018quantitative,li2015non,der2015multiscale} Traditionally, the Poisson, drift-diffusion, and Schr\"odinger equations are used to solve the problem. Solving the Schr\"odinger energy eigenvalue equation, however, consumes considerable computational resources and time. Recently, a novel simulation tool using localization landscape (LL) theory was proposed for 3D simulations, replacing the Schr\"odinger equation in solving the motion of carriers in fluctuating potentials.\cite{filoche2017localization,piccardo2017localization,li2017localization,wu2015percolation} Instead of solving the Schr\"odinger eigenvalue problem, LL theory converts the eigenvalue problem to an partial differential equation with boundaries, from which one can efficiently obtain an effective quantum potential. Use of the LL equations reduces the calculation time. In this work, we apply this method to investigate the diffusion length. Note that, due to the large simulation domain (3 $\mu$m) needed to study the lateral diffusion length ($\sim$0.5--2 $\mu$m), we only model a single QW structure sandwiched by two QBs. The electron and hole pairs are generated at the middle of the QW, such as would be realized in resonant local optical excitation or local minority carrier injection in STL, and the characteristics of diffusion are studied. The details will be discussed later.

\section{\label{sec:level2}Methodology}
\begin{figure}
\includegraphics[width=8.2cm]{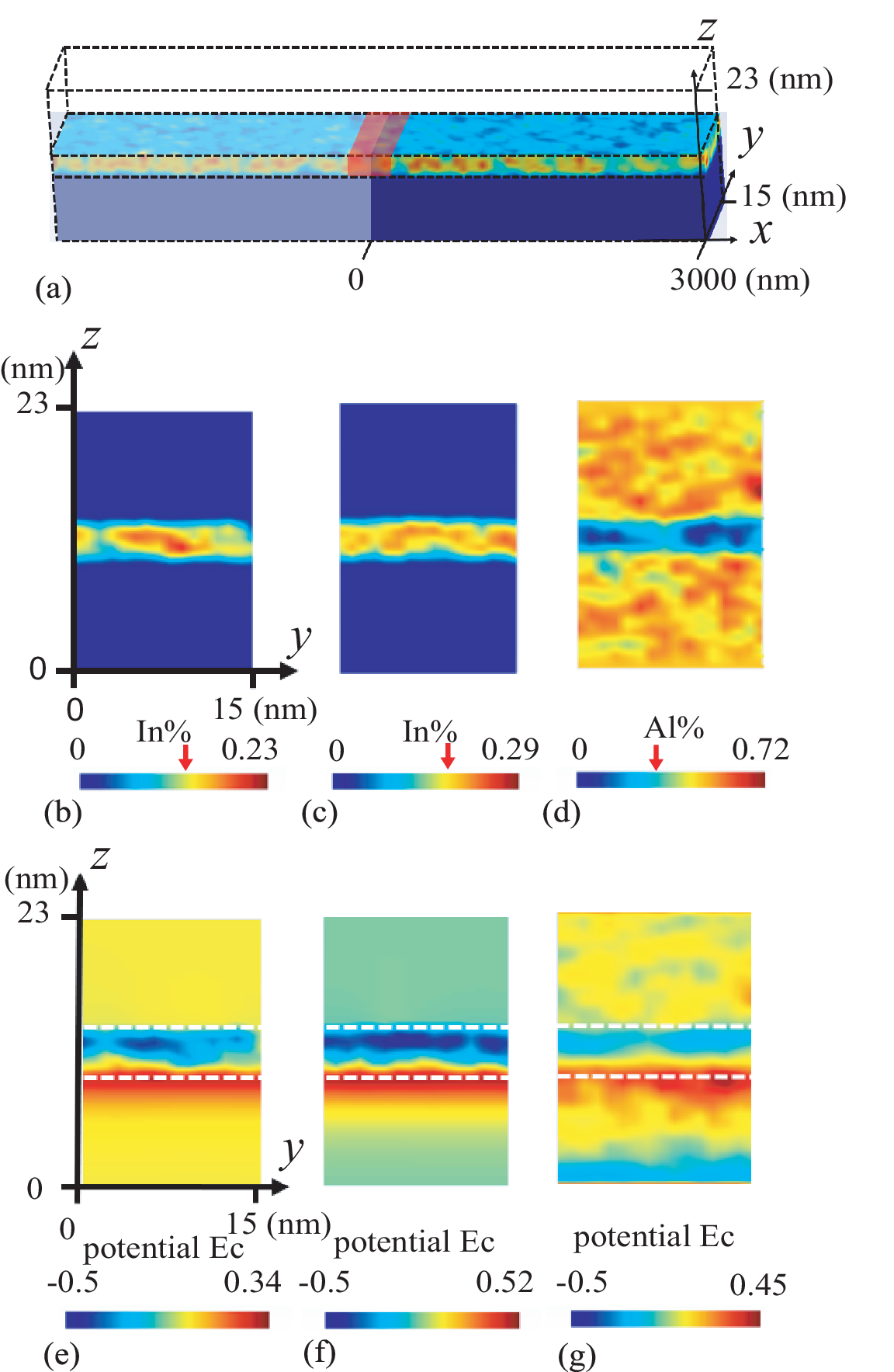}
\caption{\label{fig1} (a) Simulation structure, with dashed line representing the virtual structure. The electrons and holes are generated at the red area with a width of 10 nm. (b), (c) Indium composition maps of single QW structures for blue and green LEDs with random alloy fluctuations. (d) Single QW structure for UVC. The red arrow marks the respective average composition in the QW to the color bar. (e)--(g) Side-views of the calculated potential E$_{c}$ for blue, green, and UVC structures at x = 0, respectively. The barrier doping of this case is $2.0\times 10^{19}$ cm$^{-3}$. White dashed lines separate the QW from the QB.}
\end{figure}

As mentioned in the Introduction, modeling the 3D fluctuations in a large domain is difficult due to the high requirements for computing power and computer memory. Hence, the simulation domain size must be compromised. To understand the change of diffusion length in QWs, we design a large-area single QW measuring approximately 3 $\mu$m along the lateral $x$-direction with alloy fluctuations to mimic LED structures. The extent in the $y$-direction is limited to 15 nm due to memory constraints. The QB thickness is 10 nm and the QW width is 3 nm. The numerical complexity is depending on the mesh size and dimensionality. In our 3D modeling, the typical mesh has 4657552 nodes and 27462280 elements. The memory needed to solve this problem is about 150GB to 200GB. Moreover, the computing time is at least needs 4-8 hours with a 8 cores 16 thread Intel Xeon Silver 4110 CPU for each simulation case. Also, for each case, we need to run 15 different random seeding numbers to get the average results. To reduce the memory requirements, the excitation power (carrier generation) is imposed at one side of the simulation domain, as shown in Fig. \ref{fig1}(a), and we simulate half of the LED. The shaded volume indicates the virtual domain. The electrons and holes have different diffusion length, hence, we separate this study into n-i-n and p-i-p structures in which the holes and electrons are minority carriers, respectively. More precisely, these are n-doped QB/intrinsic-QW/n-doped QB (n-i-n) and p-doped QB/intrinsic-QW/p-doped QB (p-i-p) structures, respectively. The reason for not choosing the undoped case will be discussed later.

\begin{figure}
\includegraphics[width=8.2cm]{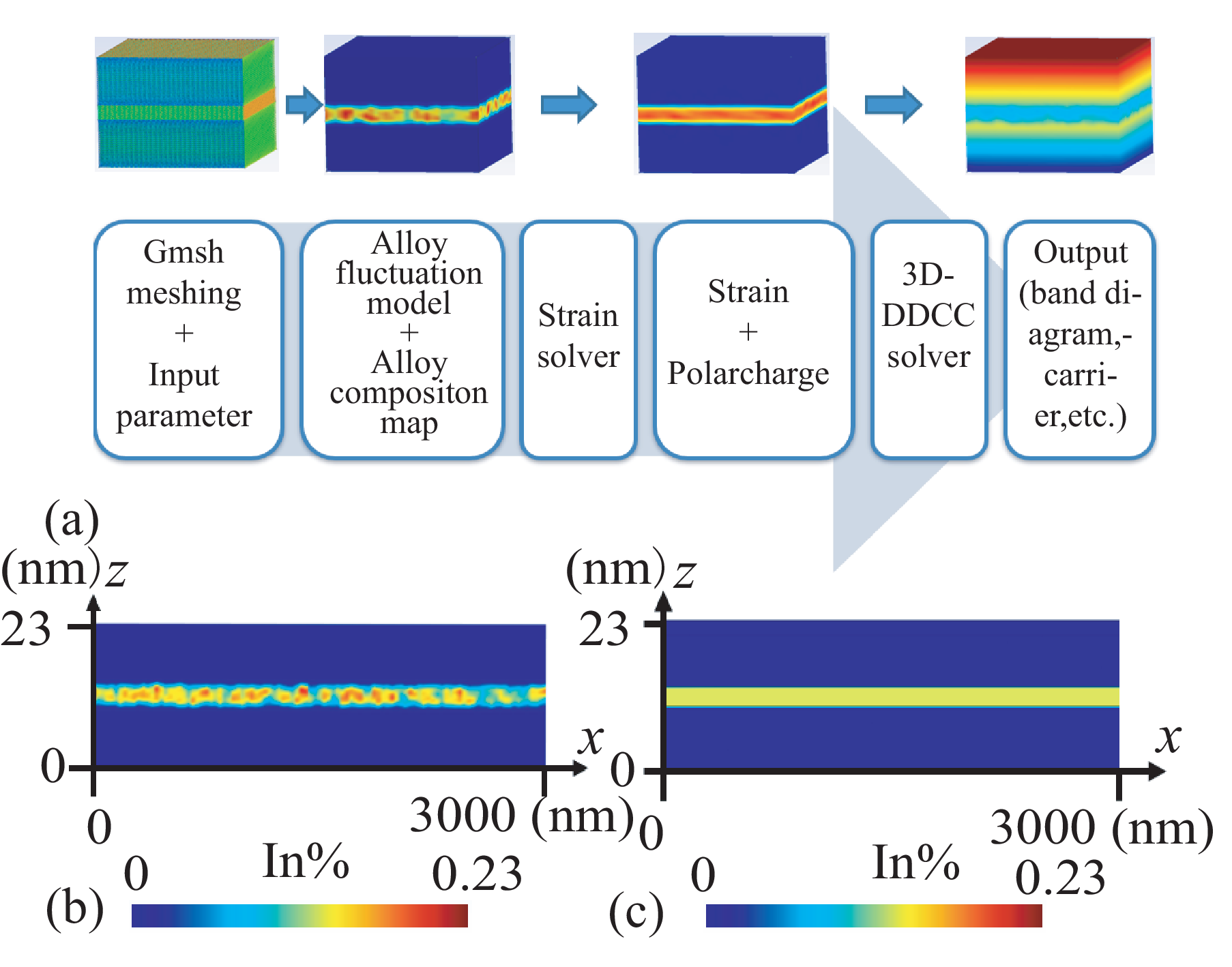}
\caption{\label{fig2} (a) Simulation flowchart. First step constructs the LED structure, then the alloy fluctuation model is used in the calculation. Finally, all parameters are input to the 3D-DDCC solver to obtain the results. The average indium composition is 15\%. (b), (c) Indium maps with and without alloy fluctuations, respectively.}
\end{figure}

To construct the structure, we first use the ``gmsh'' method\cite{geuzaine2009gmsh} to create finite-element meshes in the QW and QB regions, and set the parameters for the different layers. For the InGaN and AlGaN layers, which are the alloy materials, a random number generator selects the In or Al atoms and determines the local composition through the Gaussian averaging method. Details can be found in Refs. \onlinecite{piccardo2017localization,li2017localization}. We then use a 3D finite-element method strain solver to calculate the strain field $\varepsilon_{ij}$(r) and polarization distributions $\emph{\textbf{P}}$(r). These settings are input to the 3D Poisson solver [Eq.~(\ref{poisson1})], LL solver [Eqs.~(\ref{poisson2}) and (\ref{poisson3})], and drift-diffusion solver [Eqs.~(\ref{poisson4})--(\ref{poisson6})] to ensure that the problem is solved self-consistently. The simulation flowchart is shown in Fig. \ref{fig2}(a). 
\begin{gather}
{\nabla}({\epsilon}{\nabla}{\varphi})=e(n-p+N_{A}^{-}-N_{D}^{+}{\pm}{\rho}_{pol}),\label{poisson1}\\
\hat{H}=\frac{-\hbar^{2}}{2}\nabla{\cdot}(\frac{1}{m^{\ast}_{e,h}}\nabla)+E_{c,v},\label{poisson2}\\
\hat{H}u_{e,h}(\vec{r})=1,\label{poisson3}\\
J_{n}(r)=n{\mu}_{n}{\nabla}E_{Fn},\label{poisson4}\\
J_{p}(r)=p{\mu}_{p}{\nabla}E_{Fp},\label{poisson5}\\
{\nabla}{\cdot}J_{n,p}=-eG{\pm}e[A_{0}+B_{0}np+C_{0}(n^{2}p+np^{2})],\label{poisson6}\\
A_{0}=\frac{np-n^{2}_{i}}{{\tau}_{n}(p+n_{i}e{\frac{(E_{i}-E_{t})}{k_{B}T}})+{\tau}_{p}(n+n_{i}e{\frac{(E_{t}-E_{i})}{k_{B}T}})}.
\label{poisson7}\\
n=\int_{1/u_{e}}^{+\infty} LDOS_{3D}(E)\frac{1}{1+exp\left(\frac{(E+\Delta E_{c})-E_{fn})}{k_{B}T}\right)}\label{poisson8}\\
     p=\int_{-\infty}^{1/u_{h}} LDOS_{3D}(E)\frac{1}{1+exp\left(\frac{E_{fp}-(E+\Delta E_{v}))}{k_{B}T}\right)}\label{poisson9}\\
     LDOS_{3D}(E)=\frac{\sqrt{2}m^{\ast \frac{3}{2}}}{\pi^{2}\hbar^{3}}\sqrt{\left| E-\frac{1}{u_{e,h}} \right|}\label{poisson10}
\end{gather} 

Equation (\ref{poisson6}) is the equation of continuity, with the Shockley--Read--Hall rate A$_{0}$ given in Eq. (\ref{poisson7}). Here, $\epsilon$ is the permittivity and $\varphi$ is the electrostatic potential, E$_{Fn}$ and E$_{Fp}$ are the electron and hole quasi-Fermi levels, and n and p are the free electron and hole densities in Eqs (\ref{poisson8}) and (\ref{poisson9}), respectively, which correspond to the LDOS formula given in Ref. \onlinecite{li2017localization}. N$_{A}^{-}$ and N$_{D}^{+}$ represent the ionized acceptor and donor densities, respectively. ${\rho}_{pol}$ is the density of polarization charge. J$_{n,p}$(r), ${\mu}_{n,p}$, and ${\tau}_{n,p}$ are the current density, mobility, and NR lifetime for electrons and holes, respectively. 
For a description of where the Eqs. (\ref{poisson2}), (\ref{poisson8})--(\ref{poisson10}) originate from, please see Refs. \onlinecite{filoche2017localization,piccardo2017localization,li2017localization}.
The values for the lifetimes, radiative recombination rate B$_{0}$, mobilities, and Auger coefficient C$_{0}$ are listed in Table~\ref{tab1}, and the ionization energies are given in Table \ref{table2}. 
\begin{table}
\caption{\label{tab1} Default parameters used in the modeling. }
\begin{ruledtabular}
\begin{tabular}{c c c c}
&Blue&Green&UVC\\
\hline
Generation rate (1/cm$^{3}$s) & 1$\times$10$^{26}$& 1$\times$10$^{26}$& 1$\times$10$^{26}$ \\
Electron mobility (cm$^{2}$/Vs) & 30 &30&30\\
Hole mobility (cm$^{2}$/Vs) & 5&5&5\\
$\tau_{n},\tau_{p}$ (s) & 5$\times$10$^{-8}$& 5$\times$10$^{-8}$& 5$\times$10$^{-8}$\\
Average composition in QW& 15$\%$ & 22$\%$ &40$\%$\\
Average composition in QB& 0$\%$ & 0$\%$ &60$\%$\\
Auger coefficient (cm$^6$/s)&6$\times$10$^{-31}$&6$\times$10$^{-31}$&2$\times$10$^{-31}$\\
B$_{0}$ (cm$^3$/s)&2$\times$10$^{-11}$&2$\times$10$^{-11}$&2$\times$10$^{-11}$\\
\end{tabular}
\end{ruledtabular}
\end{table}
\begin{table}
\caption{\label{table2} Activation energies used in the modeling. }
\begin{ruledtabular}
\begin{tabular}{c c c}
& GaN & Al$_{0.6}$Ga$_{0.4}$N \\
\hline
Acceptor ionization energies (meV) & 170 & 432\\
\hline
Donor ionization energies (meV) & 25 & 76 \\
\end{tabular}
\end{ruledtabular}
\end{table}
\begin{table}
\centering
\caption{\label{table3} Materials parameters used in the modeling. }
\begin{ruledtabular}
\begin{tabular}{c c c c}
& GaN & InN & AlN \\
\hline
Permittivity ($\epsilon_r$) & 10.4 & 15.3& 10.31 \\
\hline
Effective mass ($m^{\ast}_{e}$/$m_{0}$)& 0.21 & 0.07 & 0.32\\
\hline
Effective mass ($m^{\ast}_{h}$/$m_{0}$)& 1.87 & 1.61 & 2.68\\
\end{tabular}
\end{ruledtabular}
\end{table}

There is some uncertainty about the minority carrier mobilities, as very few measurements exist: Sohi\cite{sohi2018alloy} reported 180 cm$^{2}$/Vs for majority electrons in HEMTs,  while Solowan\cite{solowan2013direct} deduced a mobility of 10 cm$^{2}$/Vs, which we interpret as the hole mobility, in undoped MQW  samples from diffusion length measurements using Einstein's relation. We therefore use conservative values of 30 and 5 cm$^{2}$/Vs for the minority electron and hole mobilities, respectively. The use of mobilities significantly lower than those for pure GaN is of course due, at least in part, to alloy disorder effects, for which there is as yet no detailed theory. We shall see below that the diffusion length varies approximately as the square root of the mobility, as expected from the simple model of the usual diffusion equation and Einstein's relation. 
For Eq. (\ref{poisson1}) in Poisson solver, the fixed boundary condition is used on the top and bottom, which is biased at 0V. For the Eq. (\ref{poisson3}), since this is based on FEM, the Neumann boundary condition was applied where $\frac{du}{dr}=0$ along surface normal direction. For drift-diffusion equation, we also set the same fixed boundary condition on top and bottom contact, where E$_{fn}$ and E$_{fp}$ are zero at the boundary.

Equations (\ref{poisson2}) and (\ref{poisson3}) are the Hamiltonians of the Schr\"odinger equations and the LL theory, and are applied to solve the effective quantum potential. The landscape functions u$_{e,h}(\vec{r})$ for electrons and holes are the solutions of Eq. (\ref{poisson3}), and 1/u$_{e,h}$ gives the effective potential incorporating the localization properties of the Schr\"odinger equation.\cite{filoche2017localization} After obtaining 1/u$_{e}$ and 1/u$_{h}$ for electrons and holes, respectively, these form the input potential for the 3D-DDCC  solver, replacing the original terms E$_{c}$ and E$_{v}$. This enables us to account for the quantum confinement effects caused by the fluctuating potential in the random alloy system. The detailed formalism and method can be found in Refs.~\onlinecite{filoche2017localization,piccardo2017localization,li2017localization}. 
The effective mass, permittivity, polar charge, and bandgap are calculated for the local In or Al composition, using linear interpolation where parameters are listed in Table \ref{table3}.

\section{\label{sec:level3}RESULTS and DISCUSSION}
Blue, green, and UVC LEDs are studied in this work.\cite{chen2018three} For the blue LED, we set the average In composition $x$ in the In$_{x}$Ga$_{1-x}$N to 15\% at the QW. For the green LED, the average In composition $x$ is 22\%. For the UVC-LED, the average Al composition $x$ in the Al$_x$Ga$_{1-x}$N composition is 40\% in the QW and 60\% in the QB. Figures \ref{fig1}(b)--\ref{fig1}(d) show the composition maps and Figs. \ref{fig1}(e)--(g) are the $x$-plane side-views of the calculated conduction band profile, E$_{c}$. In the QBs, different doping levels are applied to study the carrier diffusion length. The carriers are excited at the red region shown in Fig. \ref{fig1}(a), where we can observe carrier diffusion. The typical radiative recombination rate in the QW is about 10$^{26}$--10$^{27}$ cm$^{-3}$ s$^{-1}$. Hence, we set the generation rate in the local excitation area to 10$^{26}$ 1/cm$^{3}$s by default. When carriers are generated in the red region, they diffuse and eventually recombine in the QW. For the n-i-n structure, the majority carriers in the QW are electrons, and the holes determine the diffusion length. For the p-i-p structure, electrons determine the diffusion length because the holes already fill the whole QW. 
Note that in the p-i-p structures, without putting the excitation power (generation term) in the simulation, the hole density is already very high due to the modulation doping effect from the barriers. Hence, although the generated carriers have electrons and holes, the generated hole density is not as high as the original hole density in QW. Therefore, even the hole will still diffuse out in the simulation under excitation, the hole density does not change too much in QW compared to the original high carrier density. Therefore, the carrier diffusion length is mainly determined by the electron diffusion in the p-i-p structure. For the n-i-n structures, it would be the same concept where the diffusion length would be mainly determined by the hole diffusion. 

\subsection{Influence of doping density and polarization field on diffusion length}
\begin{figure}
\includegraphics[width=8.2cm]{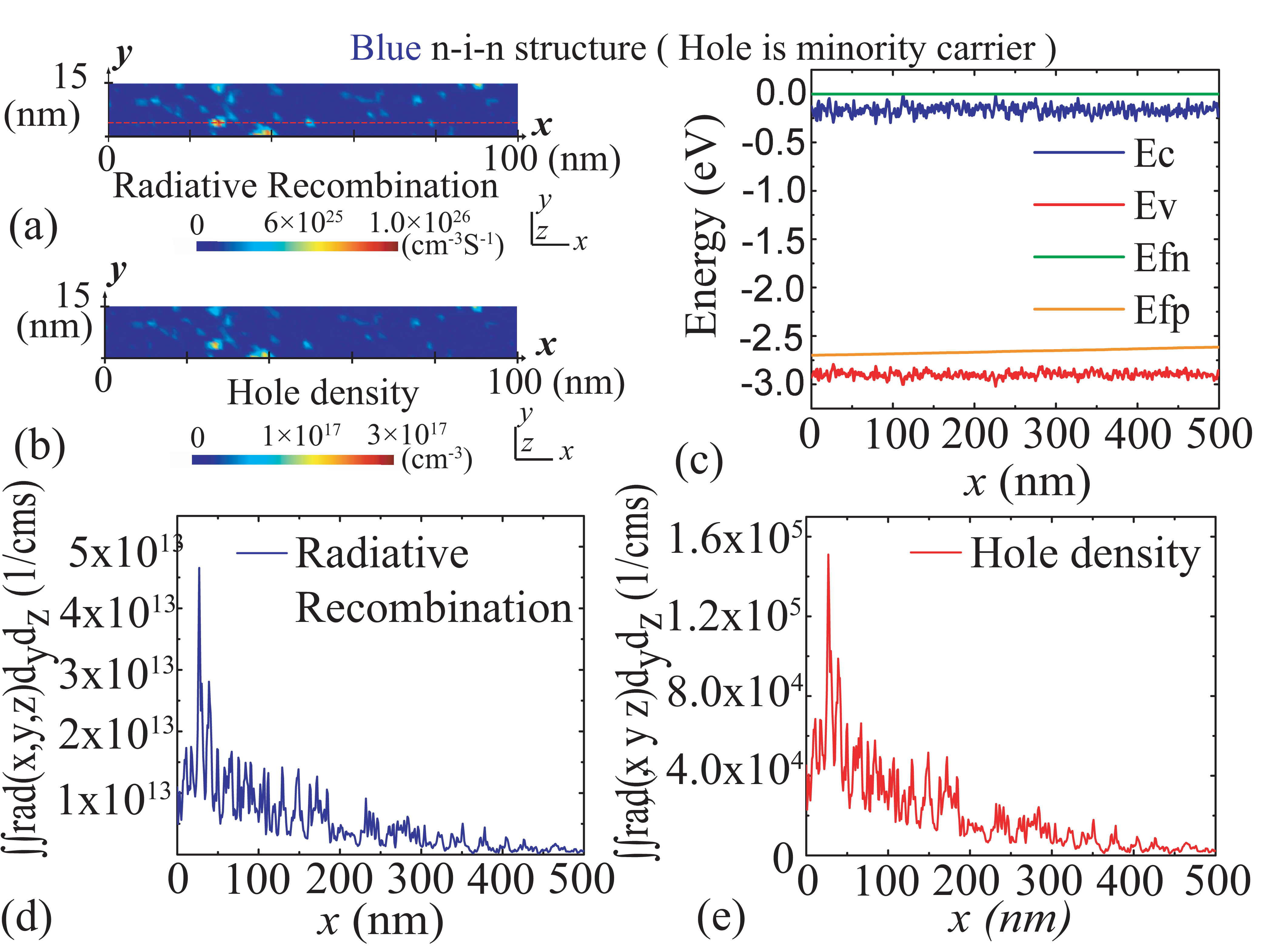}
\caption{\label{fig3} Blue LED for n-i-n structure. (See Fig. \ref{fig2})The QW in-plane view is at z = 11nm, as recombination at the 11nm plane is stronger than in the middle of the QW at 11.5nm.The QW is in the 10nm to 13nm region. (a) Radiative recombination and (b) minority carrier density (holes). (c) Band diagram affected by alloy fluctuation [along the red line in (a)], (d), (e) averaged radiative recombination and hole density along the $x$-direction, respectively.}
\end{figure}

 Figure \ref{fig3} demonstrates the typical radiative recombination rate decay\cite{danhof2012lateral,solowan2013direct,pantha2009electrical} in QWs for the blue LED in the n-i-n structure. Figure \ref{fig3}(a) shows the radiative recombination distribution within a 100-nm region. The 10-nm region from the left-hand side is the excitation area. Figure \ref{fig3}(b) shows the hole (minority carrier) density. The carriers are highly localized due to the potential fluctuations. Figure \ref{fig3}(c) shows the 1D fluctuation potential along the $x$-direction (with the red cut region over a range of 500 nm). We observe the movement of the hole quasi-Fermi level towards the center of the bandgap due to the diminishing minority hole density away from the injection region. Holes occupy localized states, where they can recombine, but this is not contradictory with hole transport: at room temperature, holes move between localized and delocalized states and conduction can occur in the latter. This part-time contribution to the conduction is reflected by the rather low mobilities observed and used in the simulations.

\begin{figure} 
\includegraphics[width=9.2cm]{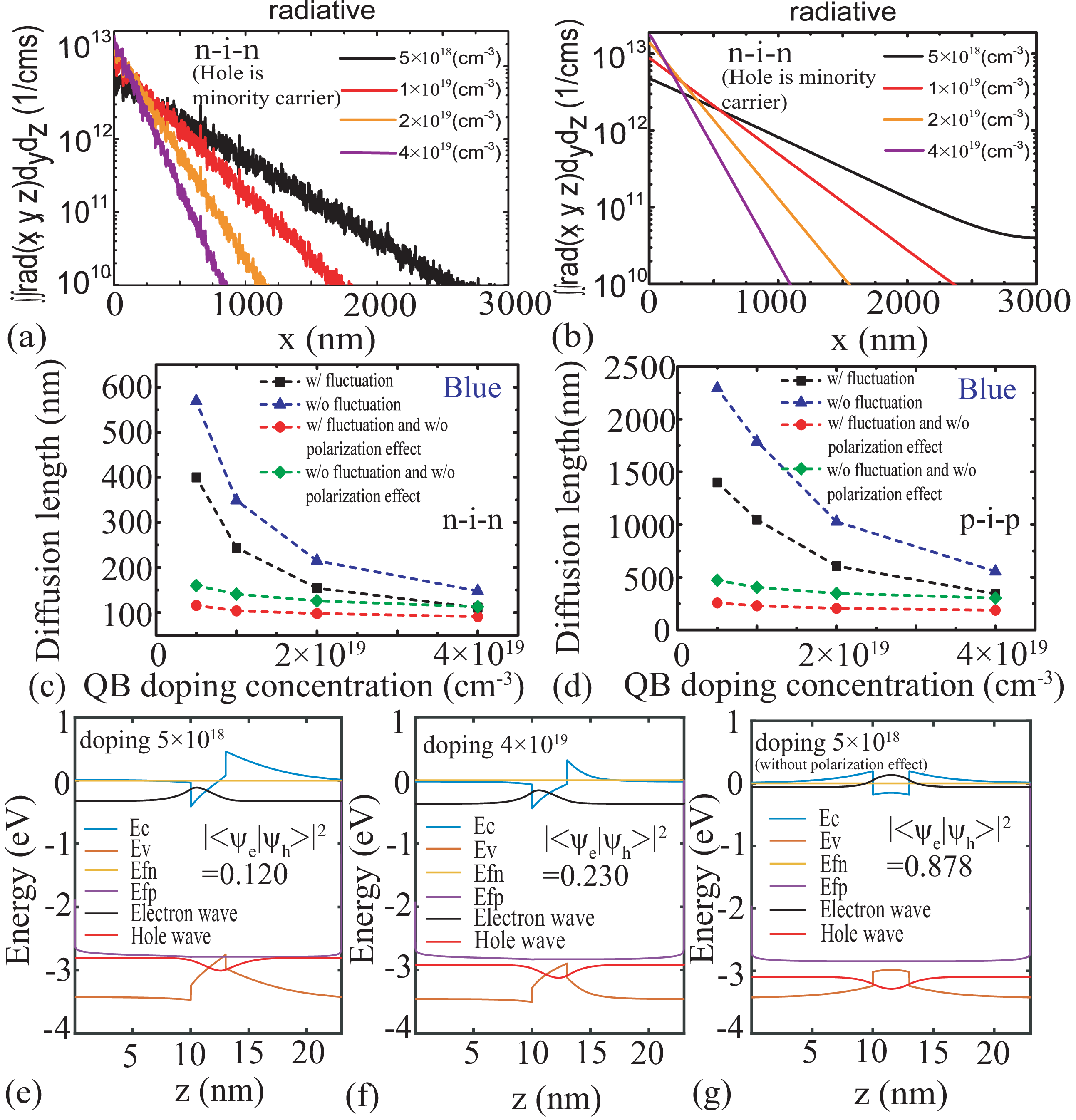}
\caption{\label{fig4} (a), (b) Averaged radiative recombination rate distribution along the lateral direction for the cases with and without random alloy fluctuations, respectively. (c) Extracted diffusion length with hole (minority) diffusion in the n-i-n structure. (d) Extracted diffusion length with electron (minority) diffusion in the p-i-p structure. The black line indicates the case with random alloy fluctuations, the blue line indicates the case without fluctuations, the red line indicates the case without polarization, and the green line indicates the case without fluctuations and without the polarization effect. (e)--(g) Electron and hole overlap for the three different situations using 1D-DDCC\cite{kawaguchi2013semipolar} simulations in the n-i-n structure.}
\end{figure}

We observe in Fig. \ref{fig3}(a) that the radiative recombination varies locally. It is hard to determine the diffusion length from the computation of a single random map. Hence, to obtain an averaged result, 15 different seeding numbers were used to generate random alloy composition maps for the simulations, and the results were averaged. Figure \ref{fig4}(a) shows the average recombination rate at each $x$-section. The shape indicates exponential decay. The diffusion length $L$ can be extracted from the exponential equation 
\begin{equation}
R(x)=Aexp\left(\frac{-(x-x_{c})}{L}\right).
\label{length}
\end{equation}
where $A$ is a fitting parameter and $L$ is the diffusion length. Figures \ref{fig4}(a) and \ref{fig4}(b) show the $y$-$z$ averaged radiative recombination rate $\emph{R(x)}$ along the $x$-direction. We now discuss the various factors that affect the fitted diffusion length.

First, the carrier diffusion lengths with and without alloy fluctuations are studied to clarify their effect on diffusion. Figure \ref{fig4} shows the result for blue n-i-n QB/QW structures with different doping densities and for the cases with and without random alloy fluctuations. Figure \ref{fig4}(a) shows that, due to the fluctuations seen in Fig. \ref{fig3}, the recombination rate $R(x)$ still fluctuates even when 15 random maps with different seeds are averaged, but is sufficiently regular to extract the diffusion length. Figure \ref{fig4}(b) is the case without alloy fluctuations, which leads to a smooth curve. Figure \ref{fig4}(c) shows the diffusion length versus doping level for cases with and without alloy fluctuations. The red line is the case with alloy fluctuations, but no polarization effects. As shown in Figs. \ref{fig4}(c) and \ref{fig4}(d), the case with alloy fluctuations has a shorter diffusion length than the case without fluctuations. The reason for these effects can be easily understood in terms of the potential fluctuations limiting the carrier diffusion in the QW: the indium composition fluctuations behave like quantum dots in the QW. The recombination in such localized sites is stronger, leading to a reduction in the diffusion length. 

However, more interestingly, the diffusion length decreases significantly as the doping density increases. This effect of the doping density is even stronger than the influence of potential fluctuations. An increase in doping density enhances the majority carrier density, but this should not affect recombination rates too much because the recombination rate is determined by the minority carrier density when the majority carrier density is high and in the degenerate condition. As the diffusion length is affected by the diffusion coefficient and the carrier lifetime, the reason for reduction of diffusion length with higher doping in QB is likely to be the change in carrier lifetime with doping, because different doping densities will not strongly affect the diffusion coefficient. The quantum-confined Stark effect (QCSE) strongly influences carrier recombination in nitride QW LEDs. Hence, we set the polarization in the simulation process to zero to investigate the influence of the QCSE. The red line shown in Fig. \ref{fig4}(c) indicates the diffusion length without polarization effects and with random alloy fluctuations: the diffusion length decreases by almost an order of magnitude without the polarization field. The p-i-p structures [see Fig. \ref{fig4}(d)] exhibit a similar trend. Moreover, the diffusion length becomes less dependent on doping density (or majority carrier density in the QW), which is to be expected as the major contributor, i.e., the change in the QCSE, is suppressed. Hence, this shows that, although random alloy fluctuations limit the diffusion length, the QCSE plays a much stronger role in the opposite direction. 

  We hereafter use a 1D simulation to demonstrate the impact of the QCSE on the radiative rate: Figs. \ref{fig4}(e)--\ref{fig4}(g) show the approximated 1D simulation\cite{kawaguchi2013semipolar} for electron and hole overlap in the QW with n = $5\times$ 10$^{18}$ cm$^{-3}$, n = $4\times$ 10$^{19}$ cm$^{-3}$, and without polarization effects. The square of the overlap integral for n = $5\times$ 10$^{18}$ cm$^{-3}$, n = $4\times$ 10$^{19}$ cm$^{-3}$, and without polarization effects increases to 0.12, 0.23, and 0.898, respectively. The reduction of the QCSE increases the square of the overlap by almost one order of magnitude, which decreases the carrier lifetime by about the same amount. With such a change, the application of the simple diffusion equation implies a factor-of-three change in the diffusion length, which is in reasonable agreement with our computations. If there is no majority carrier doping in the QW, the overlap would be further decreased and the diffusion length could even be one order of magnitude larger, extending over a few micrometers, as observed by David,\cite{david2021observation} and exceeding our simulation range. Here lies the reason for not presenting cases without doping---the long diffusion lengths that would result are beyond our current computational capabilities. This result indicates that the interpretation of carrier diffusion lengths in experimental work should carefully analyze the carrier screening conditions in each case. For the green and UVC cases, It has similar trends as the blue case. Hence we put simulation results in Fig. \ref{fig7} in Appendix.

To further investigate the influence of potential fluctuations due to random alloys, we study the influence of the doping level at different emission wavelengths for both electron and hole systems. Figure 5(a) shows the dependence of the diffusion length on the average majority carrier (electron) concentration in the QW for different doping densities (different symbols). Due to the modulation doping effect, whereby the activated carriers all transfer into the QW, the carrier density in the QW (3 nm) is much larger than the dopant density in the two 10-nm QB regions. 
\begin{figure}
\includegraphics[width=8.2cm]{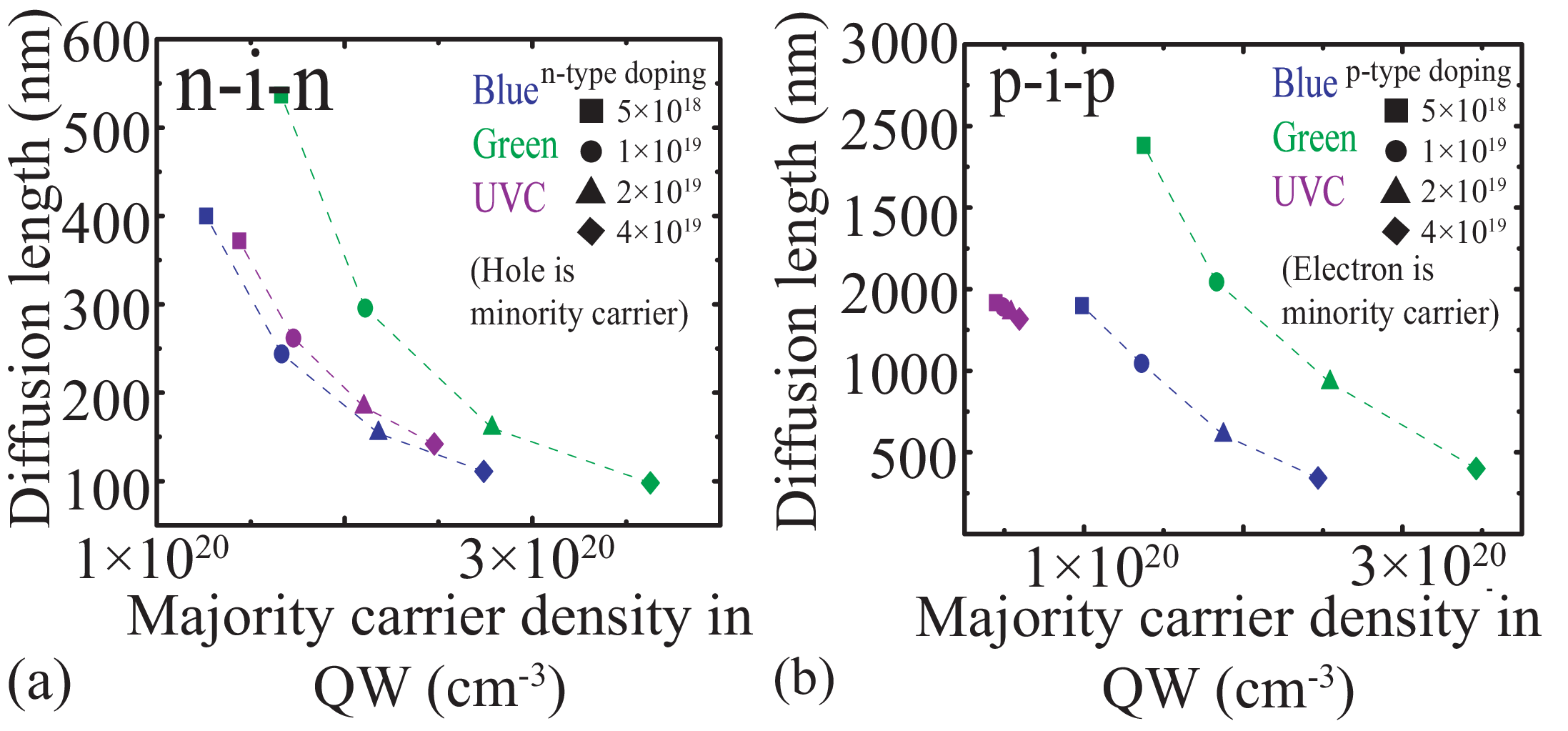}
\caption{\label{fig5} Comparison of minority carrier diffusion lengths for different doping concentrations in the QB for different structures. (a), (b) Diffusion lengths for holes and electrons versus the majority carrier density in QWs, respectively. The indium compositions for green and blue QWs are 22\% and 15\%, respectively. The average Al composition of the UVC LED is 40\% and 60\% for the QW and QB, respectively. The generation rate representing light excitation is 10$^{26}$ cm$^{-3}$s$^{-1}$. }
\end{figure} 
The results show that the diffusion length decreases as the doping density increases. As already discussed, the ionized dopants screen the field that results from fixed polarization charges (nominally at the QB/QW interface) bending in the QW (i.e., the QCSE),\cite{onuma2007quantum} and the carriers injected into the QW further screen the QCSE. The reduction in the QCSE will improve the electron and hole overlap and increase the recombination rate. Hence, the carriers will recombine more easily before they diffuse out. A more interesting result can be found in the p-i-p structures, where electrons are the minority carriers, as shown in Fig. \ref{fig5}(b). The electron diffusion length depends on the hole density in the QW in LEDs with different wavelengths. They all exhibit a similar trend, except that the UVC QW does not reach large carrier densities: when increasing the doping concentration, the majority carrier density does not increase much. The major reason for this is the high activation energy of the Mg dopant in the AlGaN layer. The activation energy of Mg acceptors for GaN is 170 meV, but is a much higher 432 meV for the Al$_{0.60}$Ga$_{0.40}$N layer. This leads to a low density of ionized acceptors in the UVC QW and a smaller carrier density in the QW. At all wavelengths, the majority carrier density in the QW appears to play a major role in limiting the diffusion length. 

\subsection{Impact of excitation power, mobility, and NR lifetime on diffusion length}
\begin{figure} 
\includegraphics[width=8.2cm]{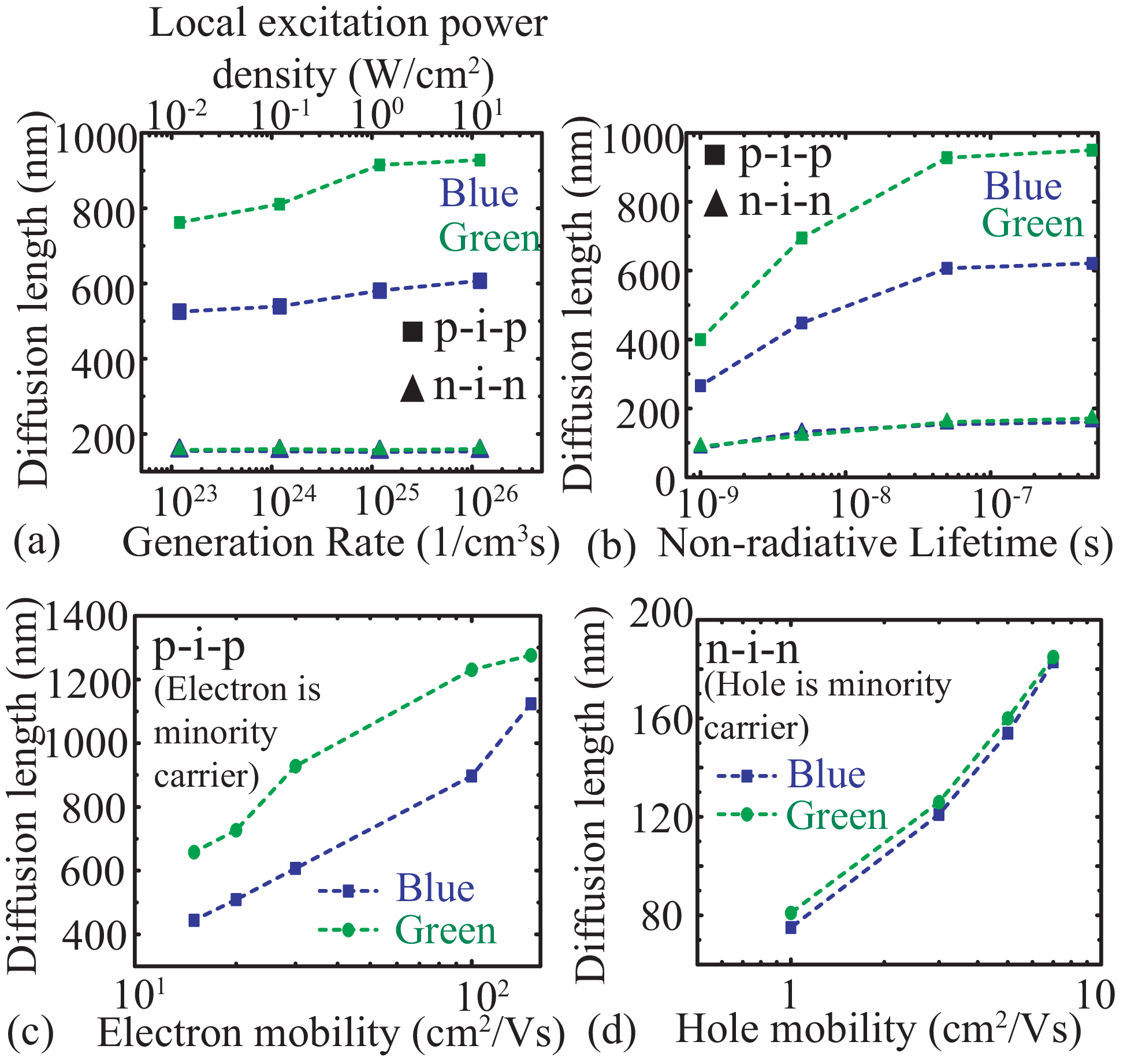}
\caption{\label{fig6} Comparison of different factors affecting the carrier diffusion length at a doping concentration of 2$\times$10$^{19}$ cm$^{-3}$ in the QB for the blue and green structures. (a) Comparison of diffusion length versus local excitation density (generation rate) in blue LEDs. (b) Influence of nonradiative lifetime on diffusion length. In green and blue n-i-n structures, the diffusion length is almost unchanged. (c), (d) Influence of mobility on diffusion length. }
\end{figure}

We now look at the different factors affecting the diffusion length when including the effect of alloy fluctuations. We simulate blue and green LED structures with 15\% and 22\% average In compositions. Due to the limited size of the simulation domain, we explore cases with diffusion lengths of less than 1 $\mu$m to avoid any boundary effects. We set the doping concentration to be $2\times 10^{19}$ cm$^{-3}$ in the QB. Figure \ref{fig6}(a) shows that, as the injected local power density increases, the electron diffusion length increases slightly. At very high excitation powers, the excess carriers cannot fully recombine locally, and so the diffusion current increases. However, this effect is not significant for the holes, possibly because of their larger effective mass. Figures \ref{fig6}(b)--\ref{fig6}(d) illustrate the influence of the nonradiative lifetime\cite{yapparov2020optimization} and carrier mobility.\cite{sohi2018alloy,sze2021physics}
Let us examine the impact of the NR lifetime and mobility in a simple model. We set the excitation power density to 10 W/cm$^{2}$. From the diffusion equation, the diffusion length $L$ is obtained as\cite{chernyak2001minority} 
\begin{equation}
L=\sqrt{D\tau},\label{L}
\end{equation}
and Einstein’s equation relates $D$ to $\mu$ as
\begin{equation}
D = \frac{k_{B}T}{e}{\mu},\label{D}
\end{equation}
where $D$ is the minority carrier diffusion coefficient, ${\tau}$ the minority carrier lifetime, k$_{B}$ is the Boltzmann constant, $T$ is the temperature, and ${\mu}$ is the minority carrier mobility. The diffusion length is proportional to the square root of the mobility and nonradiative lifetime, $\sqrt{\mu\tau}$. Figure 6(b) shows the changes in the nonradiative lifetime. With a shorter lifetime, the diffusion length decreases as the lifetime decreases. However, ${\tau}$ is not only limited by the nonradiative lifetime, but also the radiative recombination and Auger recombination. When the nonradiative lifetime is longer than 10$^{-7}$ s, the increase in diffusion length becomes saturated; this is because the radiative recombination becomes the dominant factor since B$_{0}$ is fixed at 6$\times$10$^{-11}$ cm$^3$/s. Figures \ref{fig6}(c) and \ref{fig6}(d) show the influence of the mobility parameters on the diffusion length, which broadly scales with $\sqrt{\mu}$. From the preceding discussion, we can conclude that the large differences in diffusion lengths observed in the literature can largely be attributed to differences in carrier mobility and doping density, and marginally to the excitation conditions.

\section{\label{sec:level4}CONCLUSIONS} 
In this study, we have found that alloy fluctuations effectively limit the diffusion length in nitride alloy QWs. The random composition fluctuations confine the carriers and increase the recombination rate. The mobility and nonradiative lifetime also affect the diffusion length. Higher doping concentrations increase the majority carrier, which screens the QCSE and enhances the recombination rate, thus diminishing the diffusion length. Finally, a higher local excitation power density causes the electron diffusion length to increase slightly, with the effect being even smaller in the hole case. Our results clarify the major limiting factors on the diffusion length, and differences in diffusion lengths measurements as due to different materials properties. For $\mu$LED applications or devices with a large density of defects, a shorter diffusion length is needed, and so designs with large fluctuating potentials and small QCSEs, such as semipolar structures, might be a good choice whenever possible. Sticking to c-plane structures, a good solution would be the use of field screening by fixed charge planes on both sides of the active layer as demonstrated by Young\cite{young2016germanium}. Our more recent experimental and theoretical work by Y. C. Chow et al.\cite{Chow2021} points to improvements in radiative rates due to field screening around 10 fold depending on the well width, corresponding to smaller diffusion lengths by a factor 3.

\begin{acknowledgments}
This work was supported by the Ministry of Science and Technology, Taiwan (Grant Nos. MOST 108-2628-E-002-010-MY3 and MOST 110-2923-E-002-002). The work at UCSB was supported by the National Science Foundation (Grant No. DMS-1839077), and by grants from the Simons Foundation (Grant Nos. 601952 to J.S. and 601954 to C.W.). This work was also supported by the TECCLON project, grant ANR-20-CE05-0037-01 of the French Agence Nationale de la Recherche (ANR). The data used in this paper can be requested by contacting the corresponding author. The modeling tool was developed by our laboratory, and more detailed information can be found on our website (http://yrwu-wk.ee.ntu.edu.tw/).
\end{acknowledgments}

\appendix
\section{}
\begin{figure}[h]
\includegraphics[width=8.2cm]{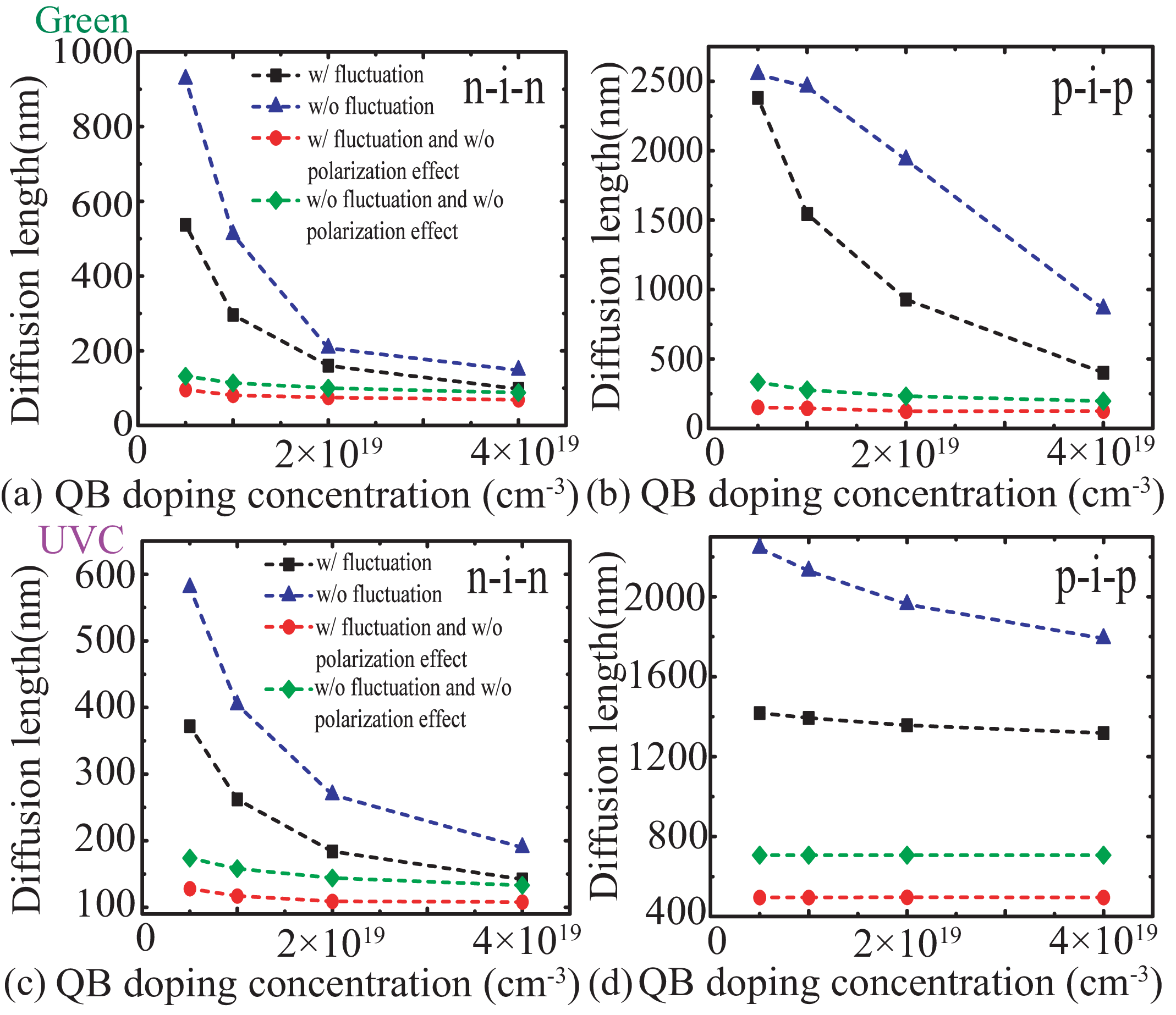}
\caption{\label{fig7} Comparison of minority diffusion lengths with and without fluctuations, without polarization effect, and with different doping concentrations in the QB. (a), (b) Green structure, (c), (d) UVC structure. The black line indicates the alloy fluctuation case, the red line indicates the case without polarization, the blue line indicates the case without fluctuations, and the green line indicates the case without fluctuations and without polarization. }
\end{figure}

Figure \ref{fig7} shows the diffusion length of green and UVC LEDs affected by alloy fluctuations, without fluctuations, and without the polarization effect. The trend is similar to that of blue LEDs when increasing the doping concentration, leading to shorter diffusion lengths. 

\nocite{*}

\bibliography{UVC_QWs_LP}

\end{document}